\documentstyle[12pt]{article}
\begin{document}
\title{
\hfill{\small TPBU-11-95}\\
\hfill{\small cond-mat/9601083}\\
\hfill{\small December 1995}\vspace*{0.5cm}\\
\sc
Upper and lower bounds on the partition function of the Hofstadter 
model
\vspace*{0.3cm}}
\author{\sc Alexander Moroz\thanks{On leave from Institute of 
Physics, Na Slovance 2, CZ-180 40 Praha 8, Czech Republic} 
\thanks{e-mail address :
{\tt am@ph.th.bham.ac.uk}}
\vspace*{0.3cm}}
\date{
\protect\normalsize
\it School of Physics and Space Research, University of Birmingham,
Edgbaston, Birmingham B15 2TT, U. K.\\
\it and\\
\it The Erwin Schr\"{o}dinger International
Institute for Mathematical Physics, \\
\it Pasteurgasse 6, A-1090 Wien, Austria}
\maketitle
\begin{center}
{\large\sc abstract}
\end{center}
 Using unitary equivalence of magnetic translation operators,
explicit upper and lower convex bounds on the partition function 
of the Hofstadter model are given for any rational ``flux" and 
any value of Bloch momenta. These bounds (i) generalize 
straightforwardly to the case of a general asymmetric hopping and 
to the case of hopping of the form $t_{jn}(S_j^n+S_j^{-n})$ with 
$n$ arbitrary integer larger than or equal $2$, and (ii) allow to 
derive bounds on the derivatives of the partition function.

\vspace*{0.3cm}


\vspace{1.2cm}

\begin{center}
{\bf (to appear in Mod. Phys. Lett. B)}
\end{center}
\thispagestyle{empty}
\baselineskip 20pt
\newpage
\setcounter{page}{1}     
\section{Introduction}
In what follows, we shall consider the Hofstadter model \cite{Ho}
described by the Hamiltonian
\begin{equation}
H = t_1 (S_1 + S_1^*) +t_2(S_2 + S_2^*),
\label{alhofham}
\end{equation}
where unitary operators $S_1$ and $S_2$, sometimes called 
magnetic translation operators, satisfy the commutation relation
\begin{equation}
S_1 S_2 = \omega S_2 S_1,
\label{s1s2rel}
\end{equation}
with 
\begin{equation}
\omega = \exp (-2\pi i\alpha),
\label{omeg}
\end{equation}
$\alpha$ being a real number. In general we shall allow for 
additional  hopping  terms of the form $t_{jn}(S_j^n+S_j^{-n})$ 
with $n$ arbitrary integer larger than or equal $2$.
However, as will become clear later, our results generalize
straightforwardly the symmetric case $t_1= t_2$
with nearest-neighbour hopping.
Therefore, for the sake of clarity, we shall firstly derive our
bounds for $t_1=t_2$ and then the relevant generalization for 
the case $t_1\neq t_2$ and additional hopping terms is 
discussed afterwords.  

The model (\ref{alhofham}) appears in many physical problems.
Originally, it appeared in the study of
the behaviour of electrons moving simultaneously in a periodic 
potential and an uniform magnetic field $B$ \cite{Ha}.
Subsequently, it has been also used in early discussions of the 
integer quantum Hall effect \cite{TKNN},
to describe superlattices and  quasicrystals \cite{Koh}, 
and  flux phases in the high-T${}_c$ superconductors \cite{Fl}.
In the case of lattice electrons moving in an uniform magnetic 
field $B$, the parameter $\alpha(\equiv \Phi/\Phi_0)$  is the 
flux per unit cell measured in units of flux quantum 
$\Phi_0 (\equiv hc/|e|)$, 
and, in the Landau gauge ${\bf A}=(-By,0,0)$,
\begin{equation}
S_1=e^{\frac{i}{\hbar}\left(\hat{p}_1+\frac{e}{c}By\right)}
=e^{\partial_1+2\pi i\alpha y},
\end{equation}
and 
\begin{equation}
S_2=e^{\frac{i}{\hbar}\hat{p}_2}=e^{\partial_2}.
\end{equation}  

In the  case where $\alpha=p/q$ is rational
with $p$ and $q$ relative prime integers,
the model is known  to enjoy $SL(2,Z)$ symmetry 
\cite{AM}, and, because of the periodicity over
an enlarged unit cell \cite{Com1}, one can define Bloch
momenta, ${\bf k}=(k_1,k_2)$, and a corresponding magnetic 
Brillouin zone \cite{Ho}.
In momentum space, the Hamiltonian $H$ and 
operators $S_1$ and $S_2$ can be written as $q\times q$ 
traceless matrices,
\begin{equation}
H(k_1,k_2) = \left(\begin{array}{cccc}
2\cos k_1 & e^{ik_2}&\ldots &e^{-ik_2}\\
e^{-ik_2}& 2\cos(k_1 + 2\pi\alpha) & \ldots & 0 \\
0 & e^{-ik_2} & \ldots & 0 \\
\vdots &\vdots &\vdots &\vdots \\
0 & 0 & \ldots & e^{ik_2}\\
e^{ik_2} & \ldots & e^{-ik_2} & 2\cos[k_1+2(q-1)\pi\alpha]
\end{array}\right),
\label{hform}
\end{equation}
\begin{equation}
S_1 = e^{ik_1}\bar{S}_1=e^{ik_1} \left(\begin{array}{ccccc}
\omega^0 & 0& \ldots &0 & 0 \\
0 & \omega^{-1} &\ldots & 0 & 0 \\
\vdots &\vdots &\vdots &\vdots &\vdots \\
0 & \ldots & 0& \omega^{-(q-2)} & 0\\
0 & \ldots & 0 & 0& \omega^{-(q-1)}
\end{array}\right),
\label{s1form}
\end{equation}
\begin{equation}
S_2 = e^{ik_2}\bar{S}_2=e^{ik_2} 
\left(\begin{array}{cccccc}
0 &  1&0&\ldots & 0 &0\\
0& 0 & 1&0& \ldots & 0 \\
0 & 0&0&1&0 & \ldots  \\
\vdots &\vdots &\vdots   &\vdots  &\vdots &\vdots \\
0 & \ldots&0 & 0 & 0& 1\\
1 & 0& \ldots & 0&  0& 0
\end{array}\right).
\label{s2form}
\end{equation} 

Despite its simple form and the extensive numerical knowledge 
available, the Hofstadter model defies exact solution 
(see Ref. \cite{WZ} for a recent progress). 
Here, we shall 
concentrate on the partition function,
\begin{equation}
Z(\beta,k_1,k_2) =\mbox{Tr}\,e^{-\beta H(k_1,k_2)},
\label{partf}
\end{equation}
where $\beta=1/T$ and $T$ is temperature.
The total partition function is then
given by the relation
\begin{equation}
Z(\beta) 
=\int_{BZ} Z(\beta,k_1,k_2) \, \frac{d^2{\bf k}}{(2\pi)^2},
\end{equation}
where the integral is over the Brillouin zone.
The calculation of the partition function $Z(\beta,k_1,k_2)$ can be 
viewed as a technical tool to calculate the density of 
states $\rho(E,k_1,k_2)$ \cite{Com2}, because $Z(\beta,k_1,k_2)$ is nothing 
but the  Laplace transform of $\rho(E,k_1,k_2)$,
\begin{equation}
Z(\beta,k_1,k_2) = \int_0^\infty e^{-\beta E}\,\rho(E,k_1,k_2)\,dE.
\label{lapl}
\end{equation}

\section{Results}
We are not yet able to calculate the partition function
$Z(\beta,k_1,k_2)$ exactly, however, we shall formulate upper and 
lower bounds on $Z(\beta,k_1,k_2)$, which hold for any rational 
$\alpha=p/q$.
To derive the lower bound on $Z(\beta,k_1,k_2)$,
we shall use the Peierls variational principle \cite{Pei} 
which states that
\begin{equation}
Z(\beta)=\mbox{Tr}\,\exp(-\beta H)\geq \sum_n 
e^{-\beta(\Phi_n,H\Phi_n)},
\label{peiv}
\end{equation}
where $\{\Phi_n\}$ is an arbitrary set of linearly
independent normalized vectors in a Hilbert space.
Eventually, to obtain the upper bound on $Z(\beta,k_1,k_2)$,
we shall use the Golden-Thompson inequality \cite{GT},
\begin{equation}
\mbox{Tr}\,e^{A+B}\leq \mbox{Tr}\left(e^Ae^B\right),
\label{gti}
\end{equation}
valid for self-adjoint operators $A$ and $B$.

Abstract inequalities  (\ref{peiv}) and  (\ref{gti})  are well
known, however, it is difficult to calculate
the traces involved and to obtain explicit analytic bounds 
in general case. In the present case,
by taking $A\equiv S_1+S_1^*$ and $B\equiv S_2+S_2^*$,
we shall calculate $e^A$, $e^B$, and the traces 
in Eqs. (\ref{peiv}) and  (\ref{gti}) explicitly 
and we shall show that
for all $\beta$, $k_1$, and $k_2$,
\begin{equation}
q\leq \mbox{max}\left(\mbox{Tr}\,e^{-\beta A},\mbox{Tr}\,
e^{-\beta B}\right)  \leq Z(\beta,k_1,k_2)\leq
\frac{1}{q}\left(\mbox{Tr}\,e^{-\beta A}\right)
\left(\mbox{Tr}\,e^{-\beta B}\right),
\label{res}
\end{equation}
where 
\begin{equation}
\mbox{Tr}\,e^{-\beta A} =\sum_{l=1}^q Q(k_1,l),\hspace*{1cm}
\mbox{Tr}\,e^{-\beta B} =\sum_{l=1}^q Q(k_2,l),
\end{equation}
and 
\begin{equation}
Q(k,l)=\exp\{-2\beta\cos[k+2\pi\alpha(l-1)]\}.
\end{equation}
Bounds (\ref{res}), which constitute the main
result of our paper, are valid for $\beta$ of both sign,
positive and negative, and they are saturated 
for $\beta=0$, i.e., at infinite temperature. The upper bound
on $Z$ in (\ref{res}) follows from (\ref{gti}) using the 
{\em clustering property} of Tr $e^A e^B=(1/q)\mbox{Tr}\,
e^A\mbox{Tr}\, e^B$
to be derived below. This clustering property is particular
to the Hofstadter model and does not depend on $\beta$
and the hopping parameters.

Both upper and lower bounds on the partition function
in (\ref{res}) are convex functions in variable $\beta$
(and hopping parameters $t_1$ and $t_2$), as the partition function 
itself.
Given convexity of $e^{-x}$, the first inequality 
(from the left) in (\ref{res}) follows from the Jensen inequality 
\cite{HLP} and the fact that $\sum_{l=1}^q \cos[k+2\pi\alpha(l-1)]=0$
for all $k$.
Using simple arguments presented in \cite{Fi}, 
bounds (\ref{res}) are used  in Sec. 4 to derive bounds 
on the derivatives  of the partition function with respect 
to $\beta$, $t_1$, and $t_2$. 

\section{Calculation of traces and derivation of bounds}
The main step in calculating traces in Eqs. (\ref{peiv}) and  
(\ref{gti}) and  deriving bounds (\ref{res}) is to use 
the {\em unitary equivalence} of operators
$\bar{S}_1$ and $\bar{S}_2$
defined by relations (\ref{s1form}) and (\ref{s2form}).
Indeed, one has  
\begin{equation}
\bar{S_2} = U \bar{S}_1 U^{-1},
\label{utrs1}
\end{equation}
where $U$ is the unitary matrix with elements given by
\begin{equation}
U_{lm}=\frac{1}{\sqrt{q}}\,\omega^{-(l-1)(m-1)}\, 
\omega \ldots \omega^{(m-1)}=
\frac{1}{\sqrt{q}}\,\omega^{-(l-1)(m-1)+m(m-1)/2}.
\label{uma}
\end{equation}
By direct calculation,
\begin{equation}
[U\bar{S}_1 U^{-1}]_{ik} = \sum_{l=1}^q
U_{il}\,U^*_{kl}\, \omega^{-(l-1)}=
\frac{1}{q} \sum_{l=1}^q \omega^{(l-1)(k-i)}\,\omega^{-(l-1)}
=\delta_{i,k-1}.
\label{utr}
\end{equation}
One can verify that $U_{l+q,m}=U_{l,m}$ and 
\begin{equation}
U_{l,m+q}=\left\{
\begin{array}{rl}
U_{l,m},&  q\hspace*{0.2cm} \mbox{odd}\\
-U_{l,m},&  q\hspace*{0.2cm} \mbox{even}.
\end{array}\right.
\end{equation}
For all $j$ and $l$ (summation over repeated indices suppressed),
\begin{equation}
U_{jl}^* U_{jl}=\frac{1}{q}\cdot
\label{jl}
\end{equation}
In general,
\begin{equation}
U_{jl}^* U_{jk} = \frac{1}{q}\,\omega^{-(j-1)(k-l)} \left[
\omega\ldots\omega^{|k-l|}\right]^{1\times\mbox{\footnotesize sgn}(k-l)}
=\frac{1}{q}\,\omega^{-(j-1)(k-l)+
\mbox{\footnotesize sgn}(k-l)\times |k-l|(|k-l|+1)/2}.
\label{jlk}
\end{equation}
The action of $U$ on the Hamiltonian $H$, $UHU^{-1}$,   
interchanges momenta $k_1$ and $k_2$ and implies that the 
spectrum of $H$ is symmetric under this interchange. It also allows
to show that the spectrum of $H$ is equivalent to the spectrum
of the Hamiltonian $\bar{H}$ of the form
\begin{equation}
\bar{H}=  U^{-1/2} D_1 U^{1/2}+ U^{1/2} D_2 U^{-1/2},
\label{funny}
\end{equation}
where $D_1$ and $D_2$ are real diagonal matrices,
$(D)_{lm}=  2\cos(k+2\pi\alpha l )\delta_{lm}$, 
with subscript labelling the dependence on either $k_1$ or $k_2$.

Due to (\ref{utrs1}),  both $\bar{S}_1$ and $\bar{S}_2$ have
{\em the same} spectrum $\Sigma$ \cite{Ex} 
which can be read off from (\ref{s1form}), i.e., 
\begin{equation}
\Sigma=\{\omega^{-j}\,|\,j=0,1,\ldots, q-1\}.
\label{s1s2spec}
\end{equation}
Normalized eigenvectors of $\bar{S}_1$ are $(\Phi_j)_l=\delta_{jl}$,
and corresponding  normalized eigenvectors of $\bar{S}_2$ are
\begin{equation}
\Psi_j=\frac{1}{\sqrt{q}}\left(
1,\omega^{-j},\omega^{-2j},\ldots,\omega^{-(q-1)j}\right).
\label{psij}
\end{equation}
One can verify that, in the standard scalar product in $C^q$, $(\Phi_j,\Phi_l)=(\Psi_j,\Psi_l)=\delta_{jl}$.
The lower bound on $Z(\beta)$ in (\ref{res}) then follows from the 
Peierls variational principle (\ref{peiv}) by taking the complete set 
of vectors to be the set of eigenvectors of either $S_1$ or $S_2$. 
Indeed, one has
\begin{equation}
(\Phi_j,H\Phi_j)=2\cos(k_1+2\pi\alpha j),
\hspace*{1cm}(\Psi_j,H\Psi_j)=2\cos(k_2+2\pi\alpha j).
\end{equation}
The same lower bound can be also obtained from
the Bogoliubov variational principle \cite{Gir},
\begin{equation}
-\beta F\geq \ln\,\mbox{Tr}\,\exp(-\beta \tilde{H})
-\beta \left\{\mbox{Tr}\, \left[
\exp(-\beta\tilde{H})(H-\tilde{H})\right]/
\mbox{Tr}\,\exp(-\beta\tilde{H})\right\},
\label{bogot}
\end{equation}
where $\tilde{H}$ is a trial Hamiltonian which is not necessarily
a dynamical operator. One uses the Bogoliubov variational principle (\ref{bogot}) with $\tilde{H}$  to be chosen either of the operators
$A$  and $B$.  In the case of general Bloch momenta $k_1$ and $k_2$, 
one has
\begin{equation}
e^A=e^{S_1+S_1^{*}} = 
\mbox{diag}\,\left\{e^{2\cos[k_1+2\pi (l-1)\alpha]}\right\},
\hspace*{0.4cm}
e^B=e^{S_2+S_2^{*}} = U \mbox{diag}\,\left\{
e^{2\cos[k_2+2\pi (l-1)\alpha]}\right\} U^{-1}.
\label{uss}
\end{equation}
The second term on the right side of Eq. (\ref{bogot}) 
then vanishes \cite{Van} and the remaining term gives
the lower bound in (\ref{res})
on the partition function $Z(\beta,k_1,k_2)$, valid for any rational
$\alpha$ and for all $\beta$,
$k_1$, and $k_2$. 

As for the upper bound (\ref{res}) on $Z(\beta,k_1,k_2)$,
using Eqs. (\ref{uss}) and (\ref{jl}) one has
\begin{eqnarray}
\mbox{Tr} \,\left( e^{-\beta A}\,e^{-\beta B}\right)&=&
\sum_{jl}
Q(k_1,l) Q(k_2,j)U_{lj}^* U_{lj}=
\frac{1}{q} \sum^q_{j,l=1} Q(k_1,l) Q(k_2,j)\nonumber\\
&=&\frac{1}{q}
\left(\mbox{Tr} \,e^{-\beta A}\right)
\left(\mbox{Tr} \,e^{-\beta B}\right),
\label{trb}
\end{eqnarray}
which is the clustering property mentioned above.
Combining Eq. (\ref{trb}) with the Golden-Thompson inequality 
(\ref{gti}) then gives the upper bound (\ref{res}) on 
$Z(\beta,k_1,k_2)$,
\begin{equation}
Z(\beta,k_1,k_2)\leq \frac{1}{q}\left(\mbox{Tr} 
\,e^{-\beta A}\right)\left(\mbox{Tr} \,e^{-\beta B}\right),
\end{equation} 
valid for any rational $\alpha=p/q$ and for all $\beta$, $k_1$,
and $k_2$.

\section{Further generalizations}
For a given $\beta$, one can check numerically that  
[after dividing (\ref{res}) by $q$] the upper and lower bounds  
on $Z(\beta,k_1,k_2)$ depend only very weakly upon $q$ and  Bloch momenta. 
The dependence of the bounds on $\beta$ is much stronger. 
As $\beta$ increases, 
\begin{equation}
\mbox{max}\left(\mbox{Tr}\,e^{-\beta A},\mbox{Tr}\,e^{-\beta B}\right)
/Z(\beta,k_1,k_2) \approx e^{-2\beta}\rightarrow 0.
\end{equation}
Therefore, the lower bound is far from being optimal and, 
in principle, it can be further improved using the Peierls 
variational principle (\ref{peiv}) with a better choice of 
the complete set of vectors than used here.
With respect to large $\beta$ behaviour, 
the upper bound (\ref{res}) on $Z(\beta,k_1,k_2)$ is much 
better because it mimicks the large $\beta$ behaviour of 
$Z(\beta,k_1,k_2)$.
The upper bound can also be further improved:
by repeatedly using inequality \cite{GT}
\begin{equation}
\mbox{Tr}\,\left(X Y\right)^{2^{m+1}}\leq
\mbox{Tr}\,\left(X^2 Y^2\right)^{2^m}
\end{equation}
for nonnegative matrices $X$ and $Y$, with $m$ integral and $\geq 0$, 
together with the Trotter formula,
\begin{equation}
e^{A+B}=\lim_{n\rightarrow\infty}
\left( e^{A/n}e^{B/n} \right)^n.
\label{trot}
\end{equation}
One finds for $2\leq l\leq m$ that
\begin{eqnarray}
\mbox{Tr}\, \left(
e^{-\beta A} e^{-\beta B}\right) 
&\geq& \mbox{Tr}\,\left( e^{-\beta A/2}e^{-\beta B/2}\right)^2
\geq  \mbox{Tr}\, \left( e^{-\beta A/l}e^{-\beta B/l}\right)^l
\nonumber\\
&\geq& \left( e^{-\beta A/m}e^{-\beta B/m}\right)^m\geq
\mbox{Tr}\,e^{-\beta H}=Z(\beta).
\label{seq}
\end{eqnarray}
Note that relation (\ref{jlk}) enables  us to calculate explicitly
\begin{equation}
\left[e^{-\beta(S_1+S_1^{-1})}\,e^{-\beta(S_2+S_2^{-1})}\right]_{lm}=
\omega^{ \mbox{\footnotesize sgn}(m-l)\times |m-l|(|m-l|+1)/2}
Q(k_1,l) \tilde{Q}(k_2,m-l),
\label{pres}
\end{equation}
where $\tilde{Q}(k_2,m-l)$ is a discrete Fourier transform,
\begin{equation}
\tilde{Q}(k_2,m-l) \equiv \frac{1}{q} \sum_j  
\omega^{-(j-1)(m-l)}\,Q(k_2,j). 
\end{equation}

Obviously, in the asymmetric case, $t_1\neq t_2$, 
and even in the case where hopping
has the form $t_{jn}(S_j^n+S_j^{-n})$ with $n\geq 2$ arbitrary 
integer, one can still use the unitary 
equivalence (\ref{utrs1}) with the matrix $U$ given 
by Eq. (\ref{uma}).
One writes $H=A+B$, where matrix $A$ ($B$) now includes 
all hopping  terms in the $x$-direction ($y$-direction)
and proceeds with matrices $A$ and $B$ as before, i.e., 
uses them as the trial Hamiltonian $\tilde{H}$ in the 
Bogoliubov variational principle (\ref{bogot}) and in the 
Golden-Thompson inequality (\ref{gti}).  Inclusion of 
hopping terms 
$t_{jn}(S_j^n+S_j^{-n})$ in the Hamiltonian $H$ does not change
the form of bounds (\ref{res}) and involves only modification of 
$Q(k,l)$. For example, in the asymmetric 
case, $t_1\neq t_2$, with the nearest-neighbour hopping, the only
change in bounds (\ref{res}) consists in replacing $A$ by $t_1 A$
and $B$ by $t_2 B$.

\section{Bounds on derivatives of $Z(\beta,k_1,k_2)$}
The partition function $Z(\beta,k_1,k_2)$ 
as a function of $\beta$ and the hopping parameters
$t_1$ and $t_2$ is a {\em convex} function.
Adapting simple arguments presented in \cite{Fi}
for a {\em concave} function, one can use
bounds (\ref{res})  to derive bounds on the derivatives 
of the partition function with respect to $\beta$,
$t_1$, and $t_2$. Indeed,
let  $f(v)$ be a convex function, i.e., $f''(v)\geq 0$.
Let us denote by $f^\pm(v)$ the upper/lower bound
on $f$ at the point $v$. Then
\begin{equation}
\frac{f^+(v_0)-f^-(v_1)}{v_0-v_1}
\leq f'(v_0) \leq
\frac{f^+(v_0)-f^-(v_2)}{v_0-v_2},
\label{derin}
\end{equation}
where $ v_1< v_0< v_2$. Inequality (\ref{derin}) then implies,
for example,
\begin{eqnarray}
\frac{1}{\beta_0-\beta_1}\left\{\frac{1}{q}
\left(\mbox{Tr}\,e^{-\beta_0 A}\right)
\left(\mbox{Tr}\,e^{-\beta_0 B}\right)
-
\mbox{max}\left(\mbox{Tr}\,e^{-\beta_1 A},\mbox{Tr}\,
e^{-\beta_1 B}\right)  \right\}
\leq 
\frac{\partial Z}{\partial\beta}(\beta_0,k_1,k_2)\nonumber\\
 \leq
\frac{1}{\beta_0-\beta_2}\left\{
\frac{1}{q}\left(\mbox{Tr}\,e^{-\beta_0 A}\right)
\left(\mbox{Tr}\,e^{-\beta_0 B}\right) -
\mbox{max}\left(\mbox{Tr}\,e^{-\beta_2 A},\mbox{Tr}\,
e^{-\beta_2 B}\right)  \right\},
\label{derpf}
\end{eqnarray}
where $\beta_1< \beta_0 <\beta_2$.

\section{Conclusions}
To conclude, we have calculated 
traces involved in general bounds on the partition 
function [Eqs. (\ref{peiv}) and (\ref{gti})] 
and established explicit upper and lower bounds on the
 partition function of the Hofstadter model
[Eq. (\ref{res})] and its derivatives
[Eq. (\ref{derpf})], valid for any rational
``flux" $\alpha$, inverse temperature $\beta$, and Bloch momenta 
$k_1$ and $k_2$. Bounds (\ref{res}) are saturated for $\beta=0$
and, in the asymmetric case, if either $t_1$ or $t_2$ approaches zero.

Bounds (\ref{res}) also imply constraints on the density 
of states of the Hofstadter model and supply constraints
imposed by the coefficients of the secular equation for 
the Hofstadter model,
\begin{equation}
\det(\lambda{\bf I}-H)=\lambda^q +a_{q-2}\lambda^{q-2}+ a_{q-4}
\lambda^{q-4} +\ldots + \det\,H = 0.
\end{equation}
Here, depending on whether $q$ is odd or even,
either $\det\,H=2[\cos(q k_1)+\cos(qk_2)]$ or
$\det\,H=4-2[\cos(q k_1)+\cos(qk_2)]$.
One can show that $a_{q-(2l+1)}\equiv 0$, $a_{q-2}=-2q$, 
\begin{eqnarray}
a_{q-4}&=&q(2q-7)-2q\cos (2\pi\alpha),\nonumber\\
a_{q-6}&=&-\frac{2}{3}q(2q^2-21 q+58) + 4q(q-6)\cos(2\pi\alpha)
-4q\cos(4\pi\alpha),
\end{eqnarray}
\cite{AT} and that except for the constant term, 
$\det\,H$, none of the coefficients $a_j$  depends on 
the Bloch momenta.

I should like to thank A. Comtet for discussions and 
suggestions at an early stage of this work, and 
R.C. Jones for careful reading of the manuscript. 
I also thank the Erwin Schr\"{o}dinger International Institute
in Wien, where part of this work was done, for its hospitality.
This work was supported  by  the UK EPSRC Grant GR/J35214
and by the Grant Agency of the Czech Republic under  
Project No. 202/93/0689.


\begin{thebibliography}{99}
\bibitem{Com1}One possible choice is, for example, a 
unit cell with $q$ lattice spacings in the $y$ direction 
and one lattice spacing in 
the $x$ direction.
\bibitem{Com2}Using transfer matrix \cite{Ho}, it can be 
shown that the spectrum of the Hamiltonian $H$ lies 
within the interval $[-4,4]$. Therefore, in order to use the formula (\ref{lapl}), one must, 
instead of $H$, consider  the Hamiltonian ${\cal H}=H+4\geq 0$.
\bibitem{Van}One can check by direct calculation that 
$\mbox{Tr}\,DB=0$ for any diagonal matrix $D$.
\end{thebibliography}
\end{document}